\documentclass{aa} % for regular A&A two-column format
\usepackage{graphicx}
\usepackage{txfonts}
\begin{document}
   \title{A multi-wavelength analysis of M81: insight on the nature of Arp's loop}

%   \subtitle{I. Overviewing the $\kappa$-mechanism}

   \author{A. Sollima\inst{1},
          A. Gil de Paz\inst{2},
	  D. Martinez-Delgado\inst{1,3},
	  R. J. Gabany\inst{4},
	  J. J. Gallego-Laborda\inst{5},
	  \and
          T. Hallas\inst{6}
          }

   \institute{Instituto de Astrofisica de Canarias, La Laguna, E38205, Spain\\
              \email{asollima@iac.es (AS)}\\
	   \and   
	      Departamento de Astrofisica, Universidad Complutense de Madrid, E28040,
	      Spain\\
          \and Max-Planck-Institut fur Astronomie, Heidelberg, Germany\\
	   \and   
	      Blackbird Observatory, New Mexico, USA\\
         \and
              Fosca Nit Observatory, Montsec Astronomical Park, Ager, Spain              
         \and
              Hallas Observatory Annex, Foresthill, USA              
             }

   \date{Received January 18, 2010; accepted March 16, 2010}

% \abstract{}{}{}{}{} 
% 5 {} token are mandatory
 
  \abstract
   {The optical ring-like structure detected by Arp (1965) around M81 (commonly
   referenced as "Arp's loop") represents one
of the most spectacular features observed in nearby galaxies. 
Arp's loop is commonly interpreted as a tail resulting from the tidal interaction between M81 and M82.
However, since its discovery the nature of this feature has remained controversial.}
   {Our primary purpose was to identify the sources of optical and infrared 
   emission observed in Arp's loop.}
   {The morphology of Arp's loop has been investigated with deep wide-field optical images.
We also measured its colors using IRAS and Spitzer-MIPS infrared images and compared them with 
those of the disk of M81 and Galactic dust
cirrus that fills the area where M81 is located.}
   {Optical images reveal that this peculiar object has a filamentary
structure characterized by many dust features overlapping M81's field. 
The ratios of far-infrared fluxes and the estimated dust-to-gas ratios 
indicate the infrared emission of Arp's loop
is dominated by the contribution of cold dust that is most likely from
Galactic cirrus.}
   {The above results suggest that the light observed at optical
wavelengths is a combination of  emission from i) a few
recent star-forming regions located close to M81, where both
bright UV complexes and peaks in the HI distribution are found, ii) the
extended disk of M81 and iii)
scattered light from the same Galactic cirrus that is responsible for the bulk
of the far-infrared emission.}


   \authorrunning
   \titlerunning

   \maketitle
%
%________________________________________________________________

\section{Introduction}
\label{intro}

The nearby galaxy M81 (NGC3031) together with the galaxies M82, NGC3077 and
NGC2976 forms one of the best local examples of a group of interacting galaxies. 
Located at a distance of $\sim$ 3.7 Mpc (Makarova et al. 2002), 
the M81 group has been a subject for many 
studies analyzing the evidence of strong interactions among its components.
The group contains remnants of tidal bridges connecting the three most prominent
galaxies, visible in neutral hydrogen (HI) survey (Gottesman \& Weliachew 1975; 
Yun et al. 1994), and over 40 dwarf galaxies (Karachentsev et al. 2001).

In particular, Arp (1965) detected an unusual ring-shaped feature in the vicinity 
of M81 while examining Schmidt photographic plates. 
This optical feature (usually referenced as "Arp's loop") is located 17 arcmin north-east of M81's
center and covers a wide area of $\sim 160$ square arcmin. It also exhibits symmetry
towards the end of the M81 disk, slightly tilted toward M82. 

The Arp's loop is commonly interpreted as a tail
resulting from the tidal interaction between M81 and M82.
However, since its discovery this interpretation has been doubted by many
authors. In fact, the region containing the M81 group of galaxies is filled
by Galactic cirrus that covers a large area of the sky near M81 (Sandage 1976).
Arp's loop exhibits colors and emission properties similar to those observed in
Galactic cirrus clouds (Abolins \& Rice 1987; Appleton et al. 1993; 
Bremnes et al. 1998), and this has raised doubts about its association with M81. 

Subsequent analyses supporting the extra-Galactic nature of Arp's loop were based
on 21cm HI observations in the M81 group region
(Yun et al. 1994). These authors detected a collimated emission in the direction
of the north-eastern part of Arp's loop which blends smoothly into the structure
and velocity of the HI disk of M81.
Numerical simulations of the system by Yun (1999) successfully reproduced 
the HI tidal debris assuming M82 and NGC 3077 approached M81 during recent
epochs.
More recently, Makarova et al. (2002) and De Mello et al. (2008) resolved the 
nebular region characterized by strong HI emission into stars using deep Hubble
Space Telescope (HST) 
observations. Their color-magnitude diagrams (CMD) showed a significant 
population of young stars (with an age of $\sim 40-160$
Myr) and an old population characterized by a well defined Red Giant Branch. 
The magnitude of the tip of the Red Giant stars assumed to be associated
with Arp's loop indicated a distance comparable with that of M81 (Karachentsev et al.
2002). These authors suggested that the old component could have been 
removed from the M81 and/or M82 disk during their mutual interaction.
Further insight was recently provided by Barker et al. (2009), who
analyzed a wide area that included M81 with deep Suprime-Cam observations.
Their results revealed no overdensity of Red
Giant stars along Arp's loop extension (which is at odds with De Mello et al.'s results), 
whereas a significant group of
young Red Supergiants and Main Sequence stars were evident in the confined region 
previously surveyed in HST studies. These authors concluded that Arp's loop was 
originally a gaseous tidal debris stream that formed stars only in the last 200-300 Myr. The same qualitative
results have also been obtained by Davidge (2009).

In this paper we present the results of a multi-wavelength analysis of Arp's
loop using the deepest optical and infrared observations available.

%__________________________________________________________________

\section{The data}

Our analysis is based on three datasets: {\it i)} a set of deep
ground-based optical images obtained at different facilities (the Fosca Nit 
Observatory and Hallas Observatory), a set of far-infrared images observed 
with the Multiband Imaging Photometer for Spitzer (MIPS) at the {\it Spitzer Space Telescope} at 24$\mu m$, 70$\mu m$
and 160$\mu m$ and with IRAS using the latest IRIS data products and {\it iii)} 
a set of high-resolution HI maps constructed from VLA interferometric
observations (from the The HI Nearby Galaxy Survey (THINGS) database; Walter et al. 2008).

The first dataset was collected with a commercially available 106mm aperture 
f/5 Takahashi FSQ refractor telescope of the Fosca Nit Observatory (FNO)
situated near Ager (Spain) at the Montsec Astronomical Park.
We used a Santa Barbara Imaging Group (SBIG) STL-11000M CCD
camera which yields a large field-of-view ($3.9^{\circ} \times 2.7^{\circ}$) at a 
plate scale of 3.5 arcsec/pixel. 
The image set consists of multiple deep exposures
through four Optec Inc. broadband (LBGR) filters. 
A set of individual 600 sec images were obtained during several
photometric nights between January and February 2008, achieving a 
total exposure time of the co-added images of 17100 sec.

High-resolution images were also gathered during four photometric nights 
between February and March 2007 with a 14.5" f/8 RCOS cassegrain
telescope situated at the Hallas Observatory Annex (HOA) situated near
Foresthill, California (USA).
A Santa Barbara Imaging Group STL-11000M camera, yielding a $41'\times27'$
field-of-view and a 1 arcsec/pixel plate scale, was attached to the Cassegrain focus of
the telescope. A set of images through four broadband (LGBR) filters were
secured and co-added, yielding a total exposure time of 51400 sec. 

The images were reduced following standard procedures for
bias correction and flat-fielding. To enhance the signal-to-noise of the faint
structures around M81, image noise was attenuated with a
Gaussian filter (Davis 1990).

\noindent

Multiband Imaging Photometer for Spitzer (MIPS) images were also retrieved from the {\it
Spitzer Infrared Nearby Galaxy Survey} (SINGS Data Release 4, Kennicut et al. 2003) public archive.
The sample consists of a set of deep images in the 24$\mu m$, 70$\mu m$ and
160$\mu m$ bands. The total exposure time is approximately 220 s, 84 s
and 25 s at 24$\mu m$,70$\mu m$ and 160$\mu m$, respectively. Individual frames
were reduced with the MIPS Instrument Team Data Analysis Tool
(Gordon et al. 2005). The background was subtracted using
the average value of an empty region of the MIPS field of view.
The uncertainties of the final absolute
calibrations were estimated at 10\% for the 24$\mu m$ data and 20\% for both the 70$\mu m$ and 160$\mu
m$ data. The combined images were then
aligned in the standard WCS reference frame. The overall field-of-view
is approximately $30' \times 44'$.

To compare the colors of Arp's loop with those of
nearby Galactic cirrus we also retrieved the newest
generation IRIS imaging products from the IRAS satellite covering a large region ($12.5^{\circ}\times12.5^{\circ}$) 
around M81 (see Fig. \ref{iras}). 
The IRIS images that we used benefit from better zodiacal-light subtraction, calibration,
zero levels, and destriping than previous versions. In
particular, the 100 $\mu$m IRIS maps represent a significant
improvement over those used by Schlegel et al. (1998).

\noindent

Neutral hydrogen maps were retrieved from the THINGS archive. The data for M81 consist in
a set of high-resolution 21cm observations performed with the VLA array of radio telescopes.
Observations were performed in B, C, and D configuration with a bandwidth of
1.56 MHz. The calibration and data reduction were performed with standard routines in the 
AIPS package. A high-resolution map was constructed adopting a 
beam size of $12.91"\times~12.41"$. The overall spatial resolution of 
the final map is 6" with a pixel size of 1.5". 
A detailed description of the data reduction procedure can be found in Walter et al.
(2008).

%__________________________________________________________________

\section{Results}
\subsection{Optical imaging}
\label{optical}
 
   \begin{figure}
   \centering
   \includegraphics[width=9cm]{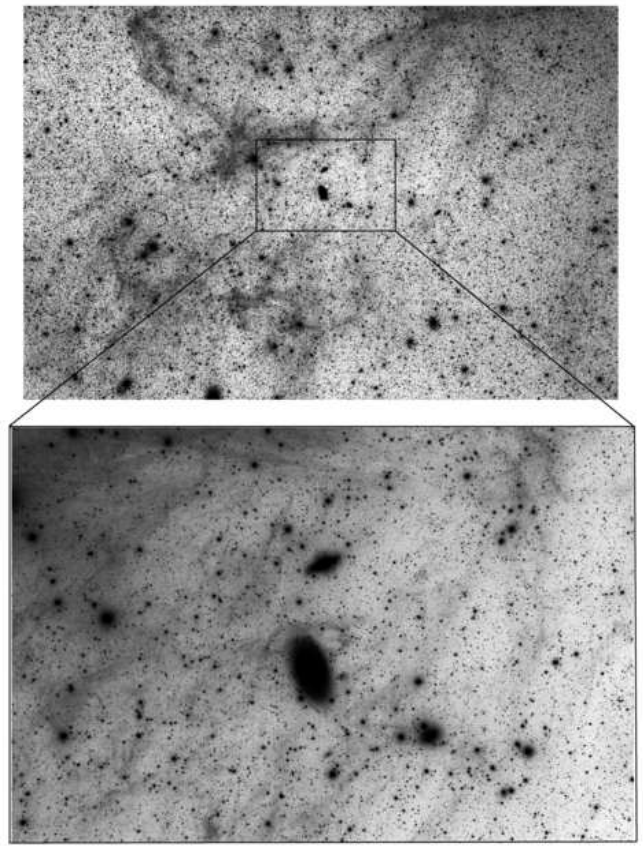}
      \caption{Upper panel: 
$2^{\circ}\times~3^{\circ}$ image of the M81 region from Witt et al. (2008; by
courtesy of S. Mandel). North is up, east to the right. 
Bottom panel: FNO image of the M81 group of galaxies.}
         \label{fsq}
   \end{figure}
 
   \begin{figure*}
   \centering
   \includegraphics[width=16cm]{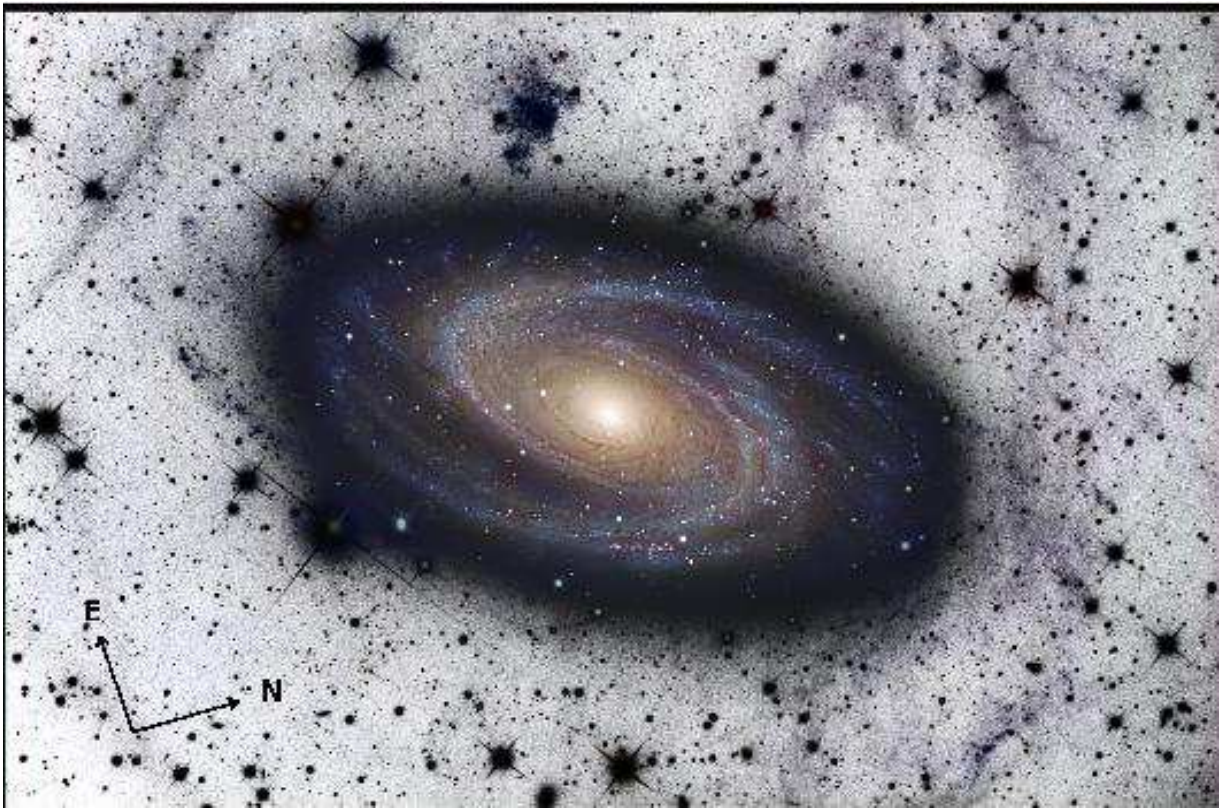}
      \caption{Combined HOA image of the M81 region. The field of view of this image is
$41'\times$27'. The north and east directions are 
indicated. Arp's loop is visible as the nebular ring crossing the disk of 
M81 on the right side of the image. Dust-absorbtion features are superimposed
over and surround the disk of M81.}
         \label{deep}
   \end{figure*}

Figure \ref{fsq} (bottom panel) shows a wide-field view around M81 obtained with the FNO telescope. 
This image clearly shows many large-scale cirrus structures. 
The width of these filaments range from $\sim$ 30 arcsec up to several arc
minutes and extend in connective patterns over several degrees (as it is clearly
visible in the panorama of this sky region displayed in the image obtained by
Witt et al. 2008; see Fig.1, top panel).

Figure \ref{deep} displays the image obtained by combining all the HOA
observations. This image covers an area of $41'\times27'$. 
This deep image reaches a point-source limiting visual magnitude of $V\sim27$.
The spectacular structure of Arp's loop is evident in the
north-eastern region. Arp's loop appears as a dim filamentary ring that wraps
and overlaps the galaxy's disk. 
A careful inspection of M81's disk reveals several dust absorption features that correspond with the
intersection of Arp's loop and the galaxy. These features,
already noticed by Arp (1965) and Efremov et al. (1986), are due to dust in the ring, indicating that part of it should be
situated between the observer and M81.
In the north-eastern part of this feature (in the region where HI strong
emission was found by Yun et al. 1994) a significant overdensity of stellar
complexes is also apparent and suggests that new stars are still forming
(Makarova et al. 2002; Mouhcine \& Ibata 2010).

\subsection{Far-infrared emission}
\label{fir_sec}

   \begin{figure*}
   \centering
   \includegraphics[width=16cm]{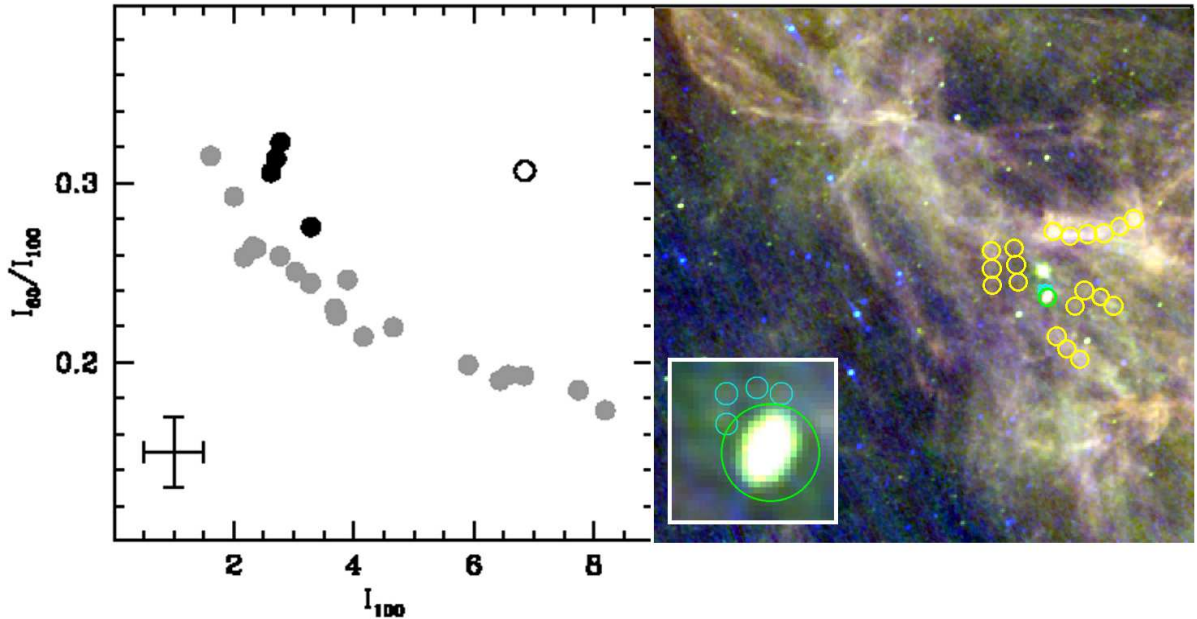}
      \caption{In the left panel the $I_{60\mu m}/I_{100\mu m}$ ratio is shown as a
function of the $I_{100\mu m}$ brightness for the loop (black filled points), M81 (open point) and the
surrounding cirrus features (grey points). The typical uncertainty of the regions along Arp's
loop is shown in the bottom part of the panel. The right panel depicts a false-color IRAS image of the M81 region. 
North is up, east is
to the left. The circular apertures used in the analysis of the Galactic cirrus (yellow
circles; 12' diameter), M81 (green circle; 12' diameter) and Arp's loop (tiny cyan circles above M81; 4'
diameter) are shown. A zoom into the M81 region is shown in the inner box.}
         \label{iras}
   \end{figure*}

As previously mentioned in Sect. \ref{intro}, far-infrared imaging is a
fundamental tool to highlight emission from dusty objects. This
is particularly important when analyzing Arp's
loop because both dust absorption in the form of dust lanes projected
against the disk of M81 and dust emission from the loop itself are
clearly seen (see Sect. \ref{optical}). The right panel of Fig. \ref{iras} 
shows an IRAS map
in a wide region ($12.5^{\circ}\times12.5^{\circ}$) around the M81 group of
galaxies constructed using images in the 12$\mu m$, 60$\mu m$ and 100$\mu m$
bands. As noted in Fig. \ref{fsq},
the surveyed area is filled with Galactic cirrus clouds. 

To compare Arp's loop with the cirrus clouds in infrared light, we measured the
fluxes for various regions of the 60$\mu m$ and 100$\mu
m$ IRAS image (see Fig. \ref{iras}). Removal of the
background (mainly of cosmic origin, because zodiacal light was previously
removed from the IRIS images) was performed using the minimum value of
the surface brightness in the $12.5^{\circ}\times12.5^{\circ}$ IRIS image for each
band. Despite its admittedly large amount of
uncertainty, this method has been already used in the past (Carey et al. 1997) 
and yields a value for the background at 100 $\mu
m$ (0.67 MJy/sr), which is very similar to the average Cosmic Infrared
Background of 0.78 MJy/sr as determined by Lagache et al. (2000).
The adopted apertures and the resulting fluxes for a number of regions in M81 and 
the Arp's loop are listed in Table 1.

In the left panel of Fig. \ref{iras} the ratio between the 60$\mu m$ and 100$\mu m$
fluxes are plotted for the main body of M81 (open black point), cirrus
clouds (grey points) and Arp's loop (black filled points) as a function of the
100$\mu m$ surface brightness. It should be noted that cirrus clouds
occupy a well defined sequence in this diagram, with high-density
cirrus showing a lower ratio than the most diffuse cirrus in the
field. This trend has been previously reported by several
authors (e.g. Abergel et al. 1994). For
comparison, the ``reference value'' commonly adopted for the $I_{60\mu
m}/I_{100\mu m}$ ratio in high-latitude cirrus is $\sim$ 0.2 (see
e.g. Arendt et al. 1998). M81 presents a high
color ($I_{60\mu m}/I_{100\mu m}\sim$ 0.3) compared to Galactic
cirrus of the same brightness, although it is not very different from 
diffuse cirrus colors or from that of the loop itself. In this
sense it is worth noting that M81 is a very quiescent object in terms
of its dust emission properties (see e.g. Dale et al. 2007). In fact, it has
the lowest $I_{60\mu m}/I_{100\mu m}$
ratio among those in the Helou (1986) sample of galaxies. More
importantly, the colors found within the circular apertures placed on the loop
(cyan circles in Fig. \ref{iras}) are not very different from the
color of the M81 disk and follow the same trend as the cirrus. This latter fact 
suggests a Galactic origin for the bulk of the far-infrared emission
associated with Arp's loop.

Despite these results, it is still conceivable that dust in
the outer disk of M81 or in a potential tidal feature around M81
could show the same colors (and follow the same surface
brightness trend) as the cirrus, because they could contain
cold dust with emission properties similar to Galactic
cirrus. The reader is referred to the recent work by Bot et al.
(2009) for an extensive discussion on the properties
of Galactic cirrus.

   \begin{figure}
   \centering
   \includegraphics[width=9cm]{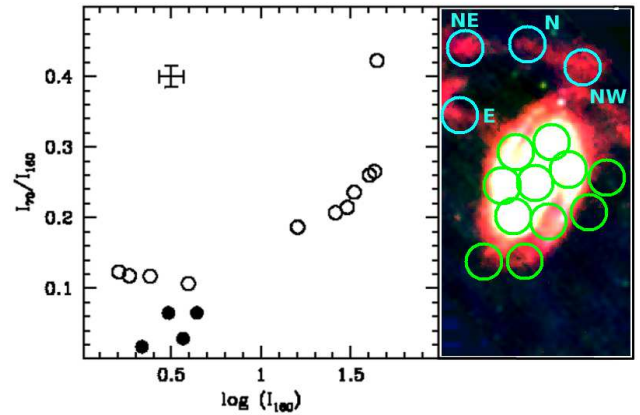}
      \caption{In the left panel the $I_{70\mu m}/I_{160\mu m}$ ratio
calculated for Arp's loop (filled circles) and the M81 disk (open circles)
are shown as a function of the $I_{160\mu m}$ brightness. The typical
uncertainty is shown in the upper part of the panel. In the right
panel the $27'\times$41' color map of the Spitzer MIPS emission for M81 was
constructed with the
24$\mu m$, 70$\mu m$ and 160$\mu m$ images. The circular apertures are also shown with the
same color code as in Fig. \ref{iras}.}
         \label{mips}
   \end{figure}
   
\noindent
 
A deeper insight into the structure of dust emission around
M81 necessarily requires deep high-resolution
(arcmin-scale) Spitzer-MIPS observations. The MIPS 160$\mu m$ data
contain information at longer wavelengths, which might offer further
clues about the possible differences (or similarities) between 
dust emission in the loop and M81. The right
panel of Fig. \ref{mips} displays a color-emission map from Spitzer. 
To construct this map, the MIPS/Spitzer 24$\mu m$ and 70$\mu m$ images were
binned and scaled to the resolution of the 160$\mu m$ image using
convolution kernels developed by Gordon et al. (2008).
To compare the far-infrared colors of Arp's loop with the stellar component of 
M81 we measured the 70$\mu m$ and 160$\mu m$ fluxes sampled in circular apertures
positioned at four regions of Arp's loop and across the M81's disk (see
the right panel of Fig. \ref{mips}). We exercised care in selecting both regions 
populated by stellar complexes and regions where only 
an extended diffuse emission is noticeable from optical images along Arp's loop. 
The measured fluxes are listed in Table 1. 
Most importantly, the 70$\mu m$/160$\mu m$ flux ratio as a function of the
160$\mu m$ brightness is shown in the left panel. 
Note that Arp's loop and the disk of M81 display very different colors. Here the 
segregation between M81 and the loop is much clearer than it is
in the IRAS $I_{60\mu m}/I_{100\mu m}$ ratio. In
particular, the ratio
70$\mu m$/160$\mu m$ appears smaller ($<$0.1 in all
cases) than that measured on
the galaxy's disk (which typically ranges between 0.2-0.4 in the inner disk and
form a plateau around 0.12 in the outer disk) along the entire
extension of the loop. The 70$\mu m$/160$\mu m$ ratios measured in the loop are
also similar
to the 60$\mu m$/160$\mu m$ ratios found by Bot et al. (2009) 
in their analysis of serendipitous observations of
Galactic cirrus contained in the SINGS Spitzer-MIPS fields.

\subsection{Comparison between far-infrared and HI emission}

\begin{figure}
   \centering
   \includegraphics[width=9cm]{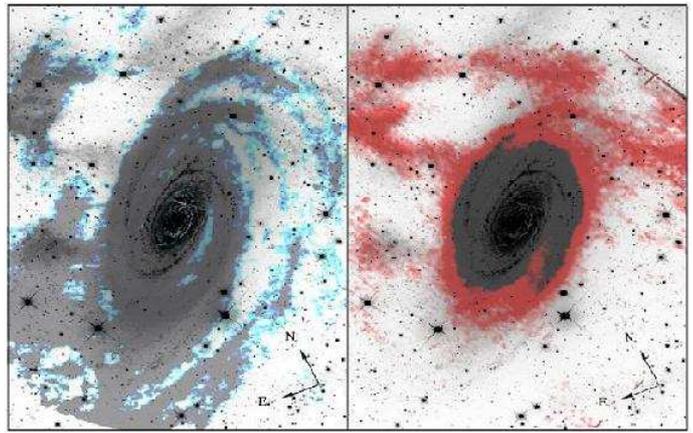}
      \caption{HOA images of the region around M81. The HI emission from Walter et al. (2008) (left panel) and the
MIPS 160$\mu m$ image (right panel) have been overplotted.}
         \label{hi}
   \end{figure}

\begin{figure}
   \centering
   \includegraphics[width=9cm]{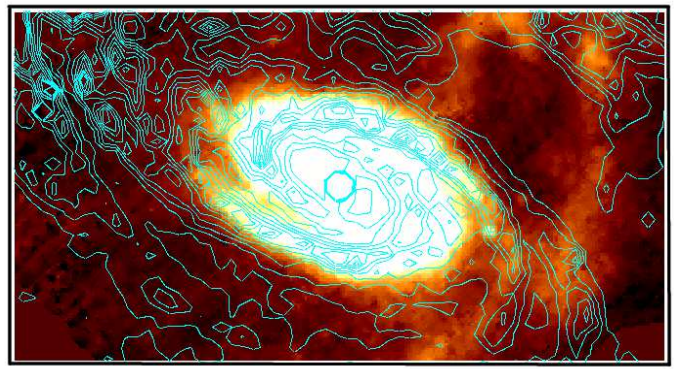}
      \caption{MIPS 160$\mu m$ image of M81. The contours of the HI emission 
from Walter et al. (2008) have been overplotted.}
         \label{contours}
   \end{figure}

Figure \ref{hi} displays the map of the HI emission measured by Walter et
al. (2008) and the MIPS 160$\mu m$ one overplotted on the optical HOA
image. Notice that the HI emission nicely follows the spiral
arms of M81. Also Holmberg IX (visible on the east of M81 in Fig.
\ref{hi}) and the southern part of Arp's loop are embedded in
clouds connected to the main body of M81. On the other hand, most of
the far-infrared emission appears to be associated to the disk of M81, with a
significant contribution from Arp's loop. 
In addition, note that the HI emission extends across the 
entire galaxy disk well beyond its optical extent, partially overlapping
Arp's loop.
It is worth noting
that while the 160$\mu m$ flux is relatively high along the entire loop
extension, the HI emission disappears in the northern part. 
To better visualize this comparison, the contours of
the HI emission measured by Walter et al. (2008) are overplotted to the MIPS
160$\mu m$ image in Fig. \ref{contours} .
The mismatch between the spatial distribution
of the HI and MIPS 160$\mu m$ emission suggests that different regions
of Arp's loop are characterized by different emitting
mechanisms. 

\begin{figure*}
   \centering
   \includegraphics[width=16cm]{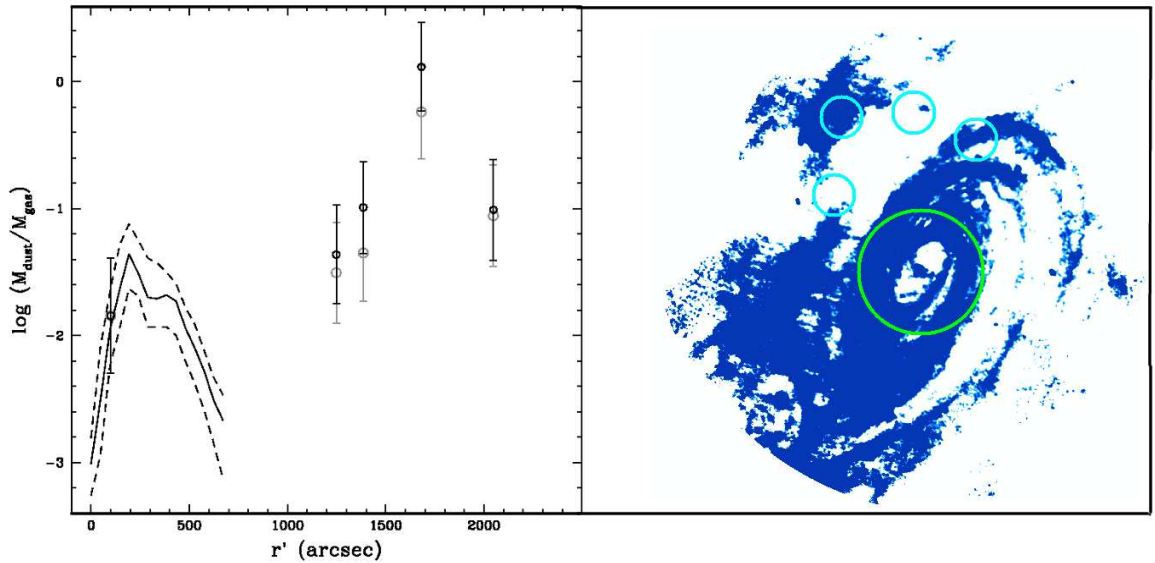}
      \caption{In the left panel the DGR calculated for Arp's loop 
      and the M81 disk are shown as a function of the equivalent angular
      distance from the galactic center. Neutral hydrogen masses were calculated using
      both the Walter et al. (2008; black points) and Yun et al. (1994; grey points) 
      HI maps. The radial DGR profile measured by
      Mu{\~n}oz-Mateos et al. (2009) is overplotted. In the right
      panel the THINGS image of M81 is shown. The circular apertures are also 
      shown with the same color code as in Fig. \ref{iras}.}
         \label{ratio}
   \end{figure*}

A more quantitative comparison was performed by estimating the relative 
fraction of dust and gas over Arp loop's extension and across the main 
body of M81. This parameter is particularly powerful in distinguishing between
the "canonical" emission of galactic disks and other dust-dominated sources.
Indeed, this quantity spans a limited range among the observed galaxies and  
has been found to show a well defined radial behavior, 
decreasing at large distances from the galactic center 
(Issa et al. 1990; Boissier et al. 2004; Mu{\~n}oz-Mateos et al. 2009).  
To estimate the total dust masses, we used MIPS fluxes at 
24$\mu m$, 70$\mu m$ and 160$\mu m$ measured in the same circular apertures
along Arp's loop and across the M81's disk defined in Sect. \ref{fir_sec} (see
also the right panel of Fig. \ref{ratio}). We adopted the relation by
Mu{\~n}oz-Mateos et al. (2009; see their Eq.A8) to convert fluxes in masses 
adopting a distance for M81 of 3.7 Mpc (Makarova et al. 2002).  
In order to obtain total gas masses, we measured the HI fluxes from the M81 intensity map
obtained by the THINGS survey (Walter et al. 2008). For this purpose the THINGS
image has been binned and scaled to the resolution of the 160$\mu m$ image
using the convolution kernels developed by Gordon et al. (2008).
The corresponding total gas masses were calculated using Eq. 3 of Walter et
al. (2008). The resulting dust and HI masses together with the corresponding 
dust-to-gas ratios (DGR) are listed in Table 1.
Equivalent radial distances from the galactic center were 
measured by assuming the position angle and axial ratio defined by  Mu{\~n}oz-Mateos
et al. (2009).
In the left panel of Fig. \ref{ratio} the obtained DGR values are plotted against
the equivalent angular radius along the semi-major axis. 
The observed profile of M81 calculated by Mu{\~n}oz-Mateos et al. (2009) is 
overplotted for comparison. 
It is evident that at distances $r'>200"$ the DGR shows the typical 
decreasing trend with galactocentric distance.
Note that the DGR values
measured along Arp's loop are significantly higher than those measured in the
central region of M81 and deviate from the general radial trend.
It is interesting to note that the maximum DGR ($log (M_{dust}/M_{gas})\sim0.12$) is 
found in the northern side of Arp's loop, where redder MIPS colors have been
measured and the HI emission drops below the detection limit.
Such a large ratio has never been observed in any galaxy analyzed by 
Mu{\~n}oz-Mateos et al. (2009) regardless of the distance from the galactic center.
On the other hand the regions along Arp's loop with smaller DGR values are those
located in the north-eastern part of the loop (where UV complexes have been
observed; De Mello et al. 2008) and in the north-western part (where the spiral
arm of M81 overlaps Arp's loop).  
The surprisingly high DGR derived along Arp's loop results from
the significantly small HI fluxes measured on the Walter et al. (2008)
maps.  To check the dependence of the obtained results on the adopted
dataset, we calso alculated the DGR using the HI map provided by Yun
et al. (1994). As apparent in Fig. \ref{ratio}, the derived DGR are
very similar to those obtained using the Walter et al.'s map. Smaller
intensities of the 21cm emission in the eastern and northern part of
Arp's loop are also evident in the single-dish map obtained by
Appleton \& van der Hulst (1988). Moreover, the HI map of Walter et
al. (2008) has been constructed using VLA data that include C and D
configurations. On the basis of all these considerations, we can exclude the
possibility that the high measured DGR could be assigned to sensitivity problems 
and/or filtering by the interferometer in the Walter et
al. (2008) data.
The high DGR measured along Arp's loop supports the idea that the 
far-infrared emission is increased due to the contribution of a
dust-dominated source. In this regard, the hypothesis about 
Galactic cirrus fits this picture because it would provide a significant
contribution to the far-infrared light, but its HI emission would be filtered out
in the VLA maps of THINGS.

\begin{table*}
\caption{Far-infrared fluxes and dust-to-gas ratios}             
\label{table:1}      
\centering          
\begin{tabular}{c c c c c l l l l l l }    
\hline\hline       
                      
 location & RA    & Dec     & diameter       & $F_{60 \mu m}^{IRIS}$ & $F_{100\mu m}^{IRIS}$ &  $F_{70 \mu m}^{MIPS}$ & $F_{160\mu m}^{MIPS}$ & $M_{dust}$       & $M_{gas}$        & $log~(M_{dust}/M_{gas})$\\
          & h m s & deg m s & \arcmin        & $Jy$                  & $Jy$                  & $Jy$                 & $Jy$                    & $10^7~M_{\odot}$ & $10^7~M_{\odot}$ &\\
\hline                    
   Arp's loop NW  & 09 54 34 & +69 16 54 & 4  & 0.962 & 3.493 & 0.285 & 4.386 & 0.393 & 9.043 & -1.362\\  
   Arp's loop N  & 09 55 43 & +69 19 28 & 4  & 0.953 & 2.953 & 0.037 & 2.171 & 2.075 &  1.586 & 0.117\\  
   Arp's loop NE & 09 57 02 & +69 18 56 & 4  & 0.903 & 2.877 & 0.106 & 3.670 & 1.390 & 14.207 & -1.009\\  
   Arp's loop E  & 09 57 09 & +69 11 21 & 4  & 0.853 & 2.788 & 0.198 & 3.048 & 0.271 & 2.6476 & -0.990\\  
   M81's disk    & 09 55 33 & +69 03 56 & 12 & 20.101 & 65.550 & 71.98 & 272.300 & 2.520 & 175.790 & -2.638\\
\hline                  
\end{tabular}
\end{table*}

%__________________________________________________________________
\section{Discussion}

We present a multi-wavelength analysis of Arp's
loop using the deepest optical and infrared images available.

Arp's loop appears to be formed by a filamentary wrap that
partially overlaps the disk of M81. The many dust-absorption features
overlapping M81's field-of-view seen in the HOA images suggest
that part of it is situated between the observer and the galaxy.

The comparison between the ratio of far-infrared fluxes emitted at
70$\mu m$ and 160$\mu m$ by Arp's loop and M81 indicates that the
emission from the loop is dominated by the contribution of cold
dust. Its far-infrared emission resembles the properties of
Galactic cirrus clouds, which also share the same colors in the IRAS bands.
The same conclusion is also supported by the analysis of the DGR measured along
Arp loop's extension, which indicates a surprisingly large relative fraction of
dust if compared with that expected at such a large distance from the galactic
center.
It is therefore likely that {\it at least part of the ring-like
structure that forms Arp's loop is constituted by Galactic
cirrus overlaying the disk of M81}. The region around M81 is indeed
known to be an area of great confusion between Galactic and
extra-Galactic objects. 
Our observations cannot exclude 
stripped material due to the interaction with M82 and NGC3077 even in
the regions of Arp's loop with small $I_{70\mu m}/I_{160\mu m}$ and high 
DGR. However, this process should have preferentially stripped dust, which
is unlikely based on the predictions of the simulations of galaxy
interactions (Yun 1999). Also in this scenario a contamination from the surrounding
Galactic cirrus would help to explain all the
observational evidences if it were overlapping Arp's loop.

On the other hand, an unambiguous distinction between these two
possibilities is very complex (see e.g. Cortese et al. 2010) and cannot be made solely on the basis of
morphological arguments or in terms of optical colors. In fact,
cirrus clouds and galaxies have largely overlapping 
colors (Bremnes et al. 1998) and emissions that encompass a wide range of wavelengths
from the UV to the far-infrared (Haikala et al. 1995).

In this context it is important to distinguish between the stellar
populations surrounding M81 and the nebular region which constitutes
Arp's loop. In recent years, many sites of recent star formation have been
reported  in M81's outer disk by GALEX (de Mello et al. 2008). 
Given the wide area covered by Arp's loop, it is possible that some of these
 star-forming regions could be accidentally located along the direct line of 
 sight path to this feature. 
Regarding this, the spatial correlation between the HI
emission and the Arp's loop provides some interesting
evidence. As previously mentioned in Sect. \ref{intro}, these HI clouds share 
similar dynamics with the disk of M81. The deep photometric analyses of this
region performed by Karachentsev et al. (2002), Makarova et al. (2002),
Sun et al. (2005), De Mello et al. (2008), Davidge (2009) and Barker et al. 
(2009) revealed evidences for
a young stellar population located at a distance comparable to M81. 
Moreover, De Mello et al. (2008) identified eight
FUV sources in this region using GALEX images. A
stellar component associated with M81 in this area therefore appears indisputable. 
However, notice that the HI emission disappears in the northern part of
the loop where both optical and infrared images show 
significant contributions. Moreover, the analysis of
CMDs obtained by Davidge (2009) and Barker et al. (2009) 
has not detected any overdensity of Red Giants along Arp's loop and only
identified a population of young stars in the confined region where the strong HI
emission was observed. Thus, although the origin of part of the optical and UV
emission in Arp's loop could be emitted by recent star formation episodes, 
most likely part of it has a Galactic origin.

It is also remarkable that both HI and GALEX emissions extend across the 
entire disk of M81 well beyond its optical cut-off, partially overlapping the 
north-western part of Arp's loop (see also Gil de Paz et al. 2007; Thilker et
al. 2007). This was also confirmed by the analysis of
HST CMDs which revealed the M81's stellar disk
population where GALEX and far-infrared emissions overlap
(Gogarten et al. 2009).

The observational evidence put forward in this paper
suggests that the structure known as Arp's loop is likely formed
by three distinct components: {\it i)} one is associated with the M81 system that
causes the emission detected in the UV and HI plus part of the
emission seen at optical wavelengths, thus dominating the morphology
of the north-eastern part of the loop, {\it ii)} a second is dominated by M81's
extended disk, which contributes to the UV and HI emission
in the north-western section of Arp's loop and, {\it iii)} another of
Galactic origin that dominates the far-infrared emission and is
responsible for the optical morphology (through scattering by dust) of the
entire ring-like structure.

\begin{acknowledgements}
This research was supported by the Spanish Ministry of Science and Innovation (MICINN) under the
grant AYA2007-65090. We thank the anonymous referee for his/her useful comments and
suggestions. We are grateful to Ignacio Trujillo, Michele
Bellazzini, Julianne Dalcanton, and Juan Carlos Mu{\~n}oz-Mateos for helpful
discussions. We also thank Steve Mandel for providing us their
wide field deep images of the M81 sky area.
\end{acknowledgements}

\end{document}